\documentclass[aps,prb,twocolumn,superscriptaddress,showpacs]{revtex4-1}

\usepackage{amssymb,amsmath}
\usepackage[sort&compress]{natbib}
\usepackage{multirow}
\usepackage{graphicx}
\usepackage{dcolumn}% Align table columns on decimal point
\usepackage{bm}% bold math

\begin{document}

\title{Magnetic Phase Transition in the Low Dimensional Compound BaMn$_{2}$Si$_{2}$O$_{7}$}

\author{J. Ma}
\affiliation{Quantum Condensed Matter Division, Oak Ridge National Laboratory, Oak Ridge, Tennessee 37831, USA}

\author{C. D. Dela Cruz}
\affiliation{Quantum Condensed Matter Division, Oak Ridge National Laboratory, Oak Ridge, Tennessee 37831, USA}

\author{Tao Hong}
\affiliation{Quantum Condensed Matter Division, Oak Ridge National Laboratory, Oak Ridge, Tennessee 37831, USA}

\author{W. Tian}
\affiliation{Quantum Condensed Matter Division, Oak Ridge National Laboratory, Oak Ridge, Tennessee 37831, USA}

\author{A. A. Aczel}
\affiliation{Quantum Condensed Matter Division, Oak Ridge National Laboratory, Oak Ridge, Tennessee 37831, USA}

\author{Songxue Chi}
\affiliation{Quantum Condensed Matter Division, Oak Ridge National Laboratory, Oak Ridge, Tennessee 37831, USA}

\author{J.-Q. Yan}
\affiliation{Materials Science and Technology Division, Oak Ridge National Laboratory, Oak Ridge, TN 37831, USA}
\affiliation{Department of Materials Science and Engineering, University of Tennessee, Knoxville, TN 37996, USA}

\author{Z. L. Dun}
\affiliation{Department of Physics and Astronomy, University of Tennessee, Knoxville, Tennessee 37996, USA}

\author{H. D. Zhou}
\affiliation{Department of Physics and Astronomy, University of Tennessee, Knoxville, Tennessee 37996, USA}

\author{M. Matsuda}
\affiliation{Quantum Condensed Matter Division, Oak Ridge National Laboratory, Oak Ridge, Tennessee 37831, USA}
\date{\today}

%%% ABSTRACT

\begin{abstract}
The structural and magnetic properties of BaMn$_{2}$Si$_{2}$O$_{7}$ have been investigated. The magnetic susceptibility and specific heat, measured using single crystals, suggest that the quasi-one-dimensional magnetism originating from the loosely coupled Mn$^{2+}$ chain carrying \emph{S}=5/2 is present at high temperatures, which is similar to the other quasi-one-dimensional barium silicates, Ba\emph{M}$_{2}$Si$_{2}$O$_{7}$ (\emph{M}: Cu and Co). The N\'{e}el temperature ($T\rm_{N}$$\sim$ 26 K) is high compared to the magnetic interaction along the chain ($J$ = -6 K). Neutron powder diffraction study has revealed that the magnetic structure is long ranged with antiferromagnetic arrangement along the chain ($c$) direction and ferromagnetic arrangement along the $a$ and $b$ axes. Detailed structural analysis suggests that the interchain interaction via Mn-O-Mn bond along the $a$ axis is relatively large, which makes the system behave more two-dimensionally in the $ac$ plane and enhances $T\rm_{N}$.
\end{abstract}

\pacs{42.50.Ct, 61.05.F-, 73.90.+f, 75.40.Cx}

\maketitle
\section{Introduction}
In low-dimensional quantum Heisenberg antiferromagnets, quantum spin fluctuations disturb long-range magnetic order. In the purely one-dimensional (1D) system, such as the $S$ = 1/2 chain system, there is no magnetic ordering even at $T$=0 K. \cite{Mikeska} However, in real materials, even a small amount of interchain coupling gives rise to three-dimensional (3D) magnetic order at a temperature much lower than the interchain coupling. \cite{Kojima} The 3D magnetic ordering is stabilized at higher temperatures in the quasi-1D Heisenberg system with larger $S$.
While the spin correlations mainly develop along the chain direction with decreasing temperature, finite interchain couplings also make the spin correlation perpendicular to the chain develop. If the chains form a layered structure, there could be an intermediate region, where the two-dimensional (2D) behavior is observed, before the system finally shows a 3D magnetic order.

The mixed alkaline earth and transition metal silicate, \emph{A}\emph{M}$_{2}$Si$_{2}$O$_{7}$ (\emph{A}: alkaline earth and \emph{M}: transition metal), is a typical low-dimensional spin system. The material has a layered structure similar to akermanites, and the presence of the large barium ions can enhance the low-dimensional character of these materials. BaCu$_{2}$Si$_{2}$O$_{7}$ is the most studied compound in this family because this is a good example of the $S$=1/2 quasi-one-dimensional antiferromagnet. In this orthorhombic $P$ n m a compound, the zigzag chain (-Cu-O-Cu-) is coupled by the corner-sharing CuO$_{4}$ plaquettes with the bond angle of Cu-O-Cu 124$^\circ$. Based on inelastic neutron scattering of BaCu$_{2}$Si$_{2}$O$_{7}$, the exchange integral values are $\mid$\emph{J}$\mid$ = 280.2 K along the chain (\emph{c} axis) and \emph{J}$_a$ = $-$5.4 K, \emph{J}$_b$ = 2.3 K, and \emph{J}$_{[110]}$ = 0.9 K in the \emph{ab} plane (between the chains). The ordering temperature is \emph{T}$_N$=9.2 K, and the ordered magnetic moment, pointing along the chain, is 0.15$\mu_B$. \cite{Tsukada, Kenzelmann} In addition, the specific one-dimensional magnetic behaviors of BaCu$_{2}$Si$_{2}$O$_{7}$, such as spin-wave modes and gaps, \cite{Kenzelmann, Hayn, Zheludev2, Zheludev3, Zheludev4} spin-flop transition, \cite{Zheludev, Glazkov, Glazkov2, Tsukada2001} and pressure effect, \cite{Rocquefelte} have been studied both experimentally and theoretically.

To understand the magnetic behavior in this system in more detail, one method is keeping the same magnetic ions, while substituting Si$^{4+}$ ions for Ge$^{4+}$, which makes the intrachain interaction enhanced.\cite{Yamada} Another option is changing the magnetic ions. In the past years, the Cu cations have been replaced by other ions with different spin values, such as Co$^{2+}$ with $S$=3/2 and Mn$^{2+}$ with $S$=5/2. For example, BaCo$_{2}$Si$_{2}$O$_{7}$ is a quasi-one-dimensional material with a lower monoclinic symmetry and a space group of $C$ 1 2/$c$ 1. In this material, the CoO$_{4}$ unit has a distorted tetrahedron structure, although the CuO$_{4}$ unit in BaCu$_{2}$Si$_{2}$O$_{7}$ has a plaquette structure. \cite{Akaki, Adams, Adams2} BaMn$_{2}$Si$_{2}$O$_{7}$ is isostructural to BaCo$_{2}$Si$_{2}$O$_{7}$. Based on the powder bulk measurements, Lu {\it et al.} found a one-dimensional behavior of the magnetism at high temperatures($J$=-7.41 K) but no long-range magnetic order down to $\sim$5 K. They also found an anomaly around 20 K in magnetic susceptibility, which was ascribed to a transition from 1D to 2D magnetism.\cite{Lu}

It is important to clarify the magnetic ground state and a possible transition from one-dimensional to two-dimensional magnetism in BaMn$_{2}$Si$_{2}$O$_{7}$. We performed magnetic susceptibility ($\chi$) and specific heat (\emph{C}$_p$) measurements using a single crystal. The $\chi$ measurement suggests a one-dimensional magnetic behavior starts around $\sim$ 26 K. Single-crystal x-ray diffraction and neutron powder diffraction measurements were also performed to determine the magnetic and lattice structures. We found that BaMn$_{2}$Si$_{2}$O$_{7}$ shows a three-dimensional magnetic order below $T\rm_{N}$$\sim$ 26 K. There was no clear sign of a dimensional crossover from 1D to 2D around $T\rm_{N}$. Structural study suggests that the interchain coupling via Mn-O-Mn bond along the $a$ axis is relatively large, which makes the system behave more two-dimensionally in the $ac$ plane and enhances $T\rm_{N}$.

\section{Experiment}
Polycrystalline BaMn$_{2}$Si$_{2}$O$_{7}$ was synthesized by the conventional solid state reaction method. A stoichiometric mixture of BaCO$_3$, MnO, and SiO$_2$ was ground together and calcined in Ar at 1200 $^\circ$C for 40 hours. A single crystal of BaMn$_{2}$Si$_{2}$O$_{7}$ was grown by the floating-zone (TSFZ) technique. The growth was carried out in flowing Ar with an IR-heated image furnace (NEC) equipped with two halogen lamps and double ellipsoidal mirrors with feed and seed rods rotating in opposite directions at 25 rpm during crystal growth at a rate of 5 mm/h. Around 5 mm size single crystal was obtained. The magnetic measurements were performed with a Superconducting Quantum Design Interference Device (SQUID) magnetometer with an applied magnetic field of $H$=2000 Oe. The \emph{M}(\emph{H}) curve was measured at 5 K. The specific heat measurements were performed with a PPMS (Physical Property Measurement System, Quantum Design) in the temperature range from 1.90 K to 300 K. Single-crystal x-ray diffraction (XRD) data were collected with a Mo \emph{K}$_\alpha$ source at room temperature.

\begin{table} [tph]
\caption{Crystallographic Data of BaMn$_2$Si$_2$O$_7$ from NPD. }
\begin{tabular}{lc}
\hline
crystal symmetry& monoclinic \\
space group& $C$ 1 2/$c$ 1 \\
\emph{a} (\AA)& 7.2761(20) \\
\emph{b} (\AA)& 12.9361(29) \\
\emph{c} (\AA)& 14.0012(33) \\
\emph{$\beta$} ($^\circ$)& 90.212(3) \\
\emph{V} (\AA$^3$) &  1317.85(53)\\
\emph{Z}  &  8 \\
density(\emph{g/cm}$^3$)  & 4.19 \\
temperature(K)  & 293 \\
$\lambda$  (\AA) & 2.4063 \\
reciprocal space (\AA$^{-1}$)& 0.387 $\le$ $Q$ $\le$ 4.731 \\
$atom$  &  $x$, $y$, $z$ \\
$\,  \,  \,  $ Ba   & 0.2049(30), 0.4717(16), 0.1260(21) \\
$\,  \,  \,  $Mn(1)& 0.7374(39)\\
$\,  \,  \,  $Mn(2)& 0.2630(40) \\
$\,  \,  \,  $Mn(3)& 0.0313(48), 0.2660(30), 0.5194(26)\\
$\,  \,  \,  $Si(1)& 0.2292(38), 0.1269(20), 0.1301(25)\\
$\,  \,  \,  $Si(2)& 0.3227(36), 0.4047(24), 0.3821(24)\\
$\,  \,  \,  $O(1)& 0.0877(32), 0.2217(15), 0.1105(18) \\
$\,  \,  \,  $O(2)& 0.1886(30), 0.3892(20), 0.2971(16) \\
$\,  \,  \,  $O(3)& 0.2340(28), 0.1310(17), 0.5127(17) \\
$\,  \,  \,  $O(4)& 0.4908(28), 0.3244(16), 0.1381(17) \\
$\,  \,  \,  $O(5)& 0.0965(28), 0.0203(17), 0.1250(21) \\
$\,  \,  \,  $O(6)& 0.2767(30), 0.1480(19), 0.2449(18) \\
$\,  \,  \,  $O(7)& 0.3818(30), 0.1088(20), 0.0478(16) \\
$R_p$  & 0.16 \\
$R_{wp}$   & 0.09\\
$\chi^{2}$ & 3.01\\
\hline
\end{tabular}
\label{lattice}
\end{table}

Neutron powder diffraction(NPD) experiments were performed at the High Flux Isotope Reactor (HFIR) of the Oak Ridge National Laboratory (ORNL). About 5 g fine powder was loaded in a vanadium-cylinder can, and mounted in a close-cycled refrigerator. The preliminary neutron diffraction data were obtained from the Wide Angle Neutron Diffractometer (WAND). High-resolution neutron powder diffraction measurements were performed using a neutron powder diffractometer, HB2A. The neutron wavelengths are $\lambda$=1.538 and 2.406 \AA\ at selected temperatures using a collimation of 12$^\prime$-open-6$^\prime$. The shorter wavelength gives a greater intensity and higher $Q$ coverage that was used to investigate the crystal structures in this low-temperature regime, while the longer wavelength gives lower $Q$ coverage and greater resolution that was important for investigating the magnetic structures of the material. The diffraction data were analyzed by the Rietveld refinement program FullProf. \cite{Juan}

The magnetic order parameter measurements were carried out at the HB1A neutron triple-axis spectrometer of HFIR. HB1A was set up in an elastic configuration with $\lambda$ = 2.359 \AA. A pyrolytic graphite (PG) (002) monochromator and analyzer were used together with collimation of 40$^\prime$-40$^\prime$-40$^\prime$-80$^\prime$. Contamination from higher order beams was removed using PG filters.

\section{Results}
\subsection{Macroscopic properties}
\subsubsection{Magnetic susceptibility}
\begin{figure}
 \centering
  \includegraphics[width=0.42\textwidth]{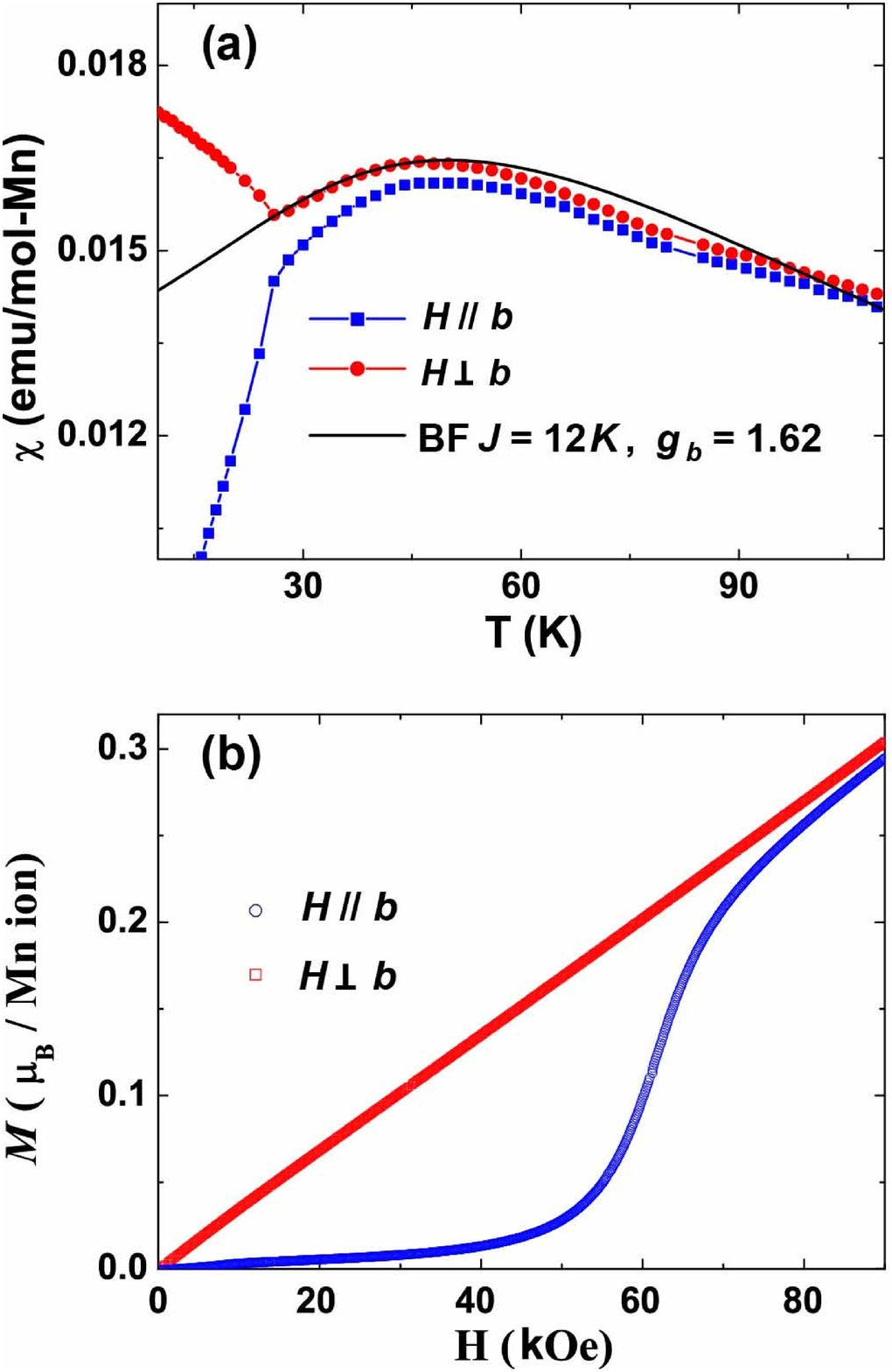}\\
\caption{(Color online) (a)The temperature dependence of the magnetic susceptibility $\chi$ along \emph{b}-axis(blue dot line) and perpendicular to the \emph{b}-axis (red dotted line) of BaMn$_{2}$Si$_{2}$O$_{7}$, which was measured with decreasing temperature from 105 \emph{K} to 10 \emph{K}. The solid black line is the fitting curve of Bonner-Fisher model. (b) Magnetization \emph{M} versus magnetic field \emph{H} along the \emph{b} axis(blue dotted line) and perpendicular to the \emph{b} axis (red dotted line) for BaMn$_{2}$Si$_{2}$O$_{7}$ up to 90 kOe at 5 \emph{K}.}
  \label{magnetism-results}
\end{figure}

The temperature dependencies of the magnetic susceptibility $\chi$ were measured parallel and perpendicular to the \emph{b}-axis with an applied field of 2000 Oe, as shown in Fig.~\ref{magnetism-results}(a). A broad symmetric peak was observed near 55 K in both directions, while the susceptibilities parallel and perpendicular to the \emph{b} axis split at 26 K, which suggests a long-range magnetic ordering with the spin easy axis along the $b$ axis. The high-temperature bump is a typical behavior of the linear-chain Heisenberg antiferromagnet and could be predicted by the modified Bonner-Fisher model for the $S$=5/2 system with two assumptions: (1) The interchain interactions are small; (2) the manganese ions interact with each other along the individual chain equally, although there are three different Mn sites and three different interactions accordingly, as described in Sec.III-B. The susceptibility $\chi$ is then expressed as, \cite{Dingle, Kim}

\begin{equation}
\chi = \frac{NS(S+1)}{3k_BT} g^2 \mu_B^2 \frac{1+u(K)}{1-u(K)} \, ,
\label{chi_m}
\end{equation}
where
\begin{equation}
\begin{split}
u(K) = {\rm coth}K - 1/K\, , \\
K = 2JS(S+1)/k_BT\, ,
\end{split}
\label{chi_m}
\end{equation}
\emph{S}, $k_B$, \emph{g}, $T$, and \emph{J} are the quantum spin number, Boltzmann constant, Lande factor, temperature, and antiferromagnetic interaction exchange along the chain, respectively.
\begin{equation}
H = - 2J \sum_{i, j} \bf{S_i} \cdot \bf{S_j} \, ,
\label{Heisen}
\end{equation}

The quantitative fitting curve (black line) is compared to the experiential data (red/blue dots) in Fig~\ref{magnetism-results}(a). In the fitting \emph{S} was fixed at 5/2. \emph{g}$_b$ was fitted to be 1.62, which is smaller than 2, suggesting that the $g$-factor is anisotropic due to the spin-orbit coupling in this Mn system. In addition, the AF interaction along the chain \emph{J} is -6.0(0.2) K (-0.51(0.2) meV). Although it is much smaller than the value of BaCu$_{2}$Si$_{2}$O$_{7}$ (\emph{J}$\rm_{Cu}$ $\approx$ -280 K = -24.1 meV), it is comparable to the other one-dimensional manganese systems, such as, [(CH$_3$)$_4$N][MnCl$_3$], which has an exchange energy -6.4 K with the sharing faces of the octahedrally coordinated manganese atoms\cite{deJongh}.

Figure~\ref{magnetism-results}(b) shows the magnetization, \emph{M}, as a function of applied magnetic field longitudinal and transverse to the \emph{b} axis at 5 K. It reveals that a spin flop transition occurs at $\sim$5.5 T when the field is applied along the easy axis ($b$ axis).

\subsubsection{Specific heat}

\begin{figure}
 \centering
  \includegraphics[width=0.48\textwidth]{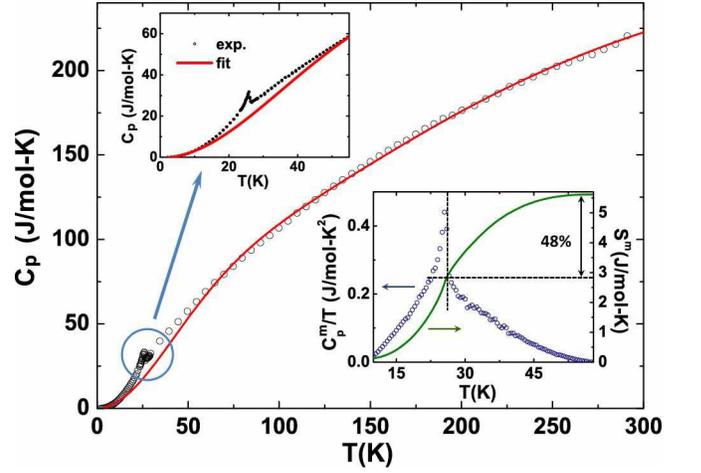}\\
\caption{(Color online) Temperature dependence of specific heat for BaMn$_{2}$Si$_{2}$O$_{7}$. The red curve shows the fit of lattice specific heat with Thirring model. Upper inset presents the enlarged temperature range(0 $\le$ $T$ $\le$ 55 K). Lower inset presents the $Cp^m$/$T$ versus $T$ and the numerical integration of the entropy S after subtracting the lattice specific heat around $T\rm_{N}$.}
  \label{heat_capacity}
\end{figure}

Figure~\ref{heat_capacity} shows the specific heat in BaMn$_{2}$Si$_{2}$O$_{7}$. A sharp peak was observed at 26 K, consistent with $T\rm_{N}$ observed in the magnetic susceptibility measurement. The sharp peak was not observed in the previous measurement with a powder sample. \cite{Lu} The magnetic specific heat was obtained by subtracting the lattice contribution. The Thirring model was applied to estimate the lattice specific heat,\cite{Thirring, Yan}
\begin{equation}
  C_{lattice}  = 3 N R ( 1 + \sum_{n=1}^\infty b_n \mu^{-n}) \, ,
\label{Thirring_Cp}
\end{equation}
where $N$ is the number of atoms in the unit cell, $R$ is the ideal gas constant, $\mu$ = (2$\pi$$T$ / $\theta_D$)$^2$ + 1, and $\theta_D$ is the Debye
temperature. In the fitting, $b_1$ = - 2.17(3), $b_2$ = 2.87(5), $b_3$ = 1.70(3), and $\theta_D$ = 749(8) K. A reasonable accuracy is obtained as demonstrated by the red solid line in Fig.~\ref{heat_capacity}. The upper inset of Fig.~\ref{heat_capacity} shows the enlarged comparison between the experimental data and the lattice fitting around $T\rm_{N}$. By subtracting the lattice specific heat, we obtained the entropy change involved in the transition from the area under the anomaly by a $C_p$/T versus $T$ plot.

The lower inset of Fig.~\ref{heat_capacity} presents the magnetic $C_p$/T and the entropy S after subtracting the lattice specific heat around $T\rm_{N}$.
A broad maximum is observed around 30 K, in addition to the sharp magnetic peak at $\sim$26 K. The similar broad peak was also observed in the previous study with a powder sample.\cite{Lu} The entropy change takes place over a wider temperature range from 55 K to 15 K, and $\sim$ 48$\%$ is entropy lost at $T\rm_{N}$ due to the short-range ordered state in the low-dimensional magnet.

\subsection{Structural properties}
\subsubsection{Crystallographic structure}

\begin{figure}
 \centering
  \includegraphics[width=0.38\textwidth]{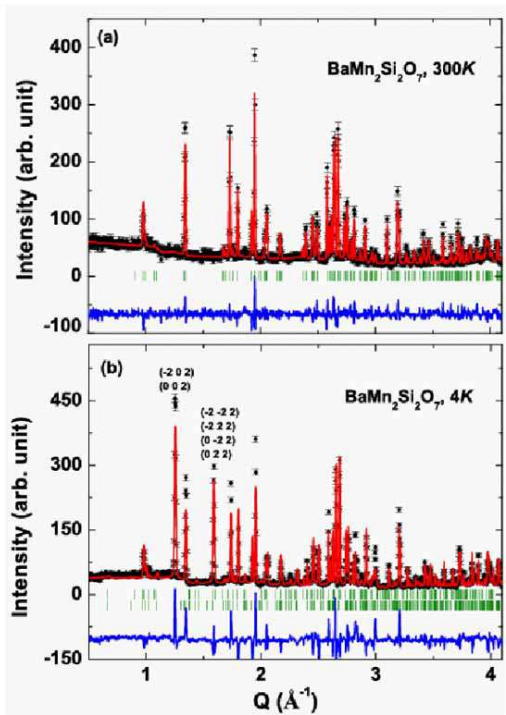}\\
\caption{(Color online) Neutron diffraction patterns (markers) and Rietveld refinements (red line) for BaMn$_{2}$Si$_{2}$O$_{7}$ at 300 K (a) and 4 K (b). Blue curves at the bottom of each panel are difference curves. The green tick marks in (a) are the reflection positions for the nuclear phases, while the green upper and lower tick marks in (b) are the reflection positions for the nuclear and magnetic phases, respectively. The first magnetic phases are labeled. The refinements were done with the \emph{C} 1 2\emph{/c} 1 space group.}
  \label{neutron_diff}
\end{figure}

The crystal structure was obtained from the refinement of the 300 K data of HB2A shown in Fig.~\ref{neutron_diff}(a). The NPD pattern can be indexed in a monoclinic unit cell with \emph{a} = 7.2767(16) \AA, \emph{b} = 12.9370(27) \AA, \emph{c} = 14.0022(31) \AA, and $\beta$ = 90.213(4)$^\circ$. The space group is \emph{C}1 2\emph{/c} 1 with one manganese atom at the 8\emph{f} ($x$, $y$, $z$) site and two manganese atoms at the 4\emph{e} (0, $y$, 1/4) site, the other atoms (barium, silicon, and oxygen) are all located at the 8\emph{f} site. Detailed information about the structural refinement and the atomic coordinates is summarized in Table ~\ref{lattice}.

The single-crystal XRD was also applied to determine the nuclear structure, and it is consistent with that determined by the NPD measurement. Figure~\ref{structure} illustrates the refined lattice structure and the Mn- nteractions in the \emph{ac} plane.

\subsubsection{Magnetic structure}
Below $T\rm_{N}$, magnetic Bragg reflections were observed. Figure.~\ref{neutron_diff}(b) shows the low-temperature NPD pattern of HB2A with the refinement results of both magnetic and nuclear structure unit cell. The diffraction patterns with good statistics were also collected at several temperatures. No observable structural change was detected around $T\rm_{N}$.

\begin{figure}
 \centering
   \includegraphics[width=0.5\textwidth]{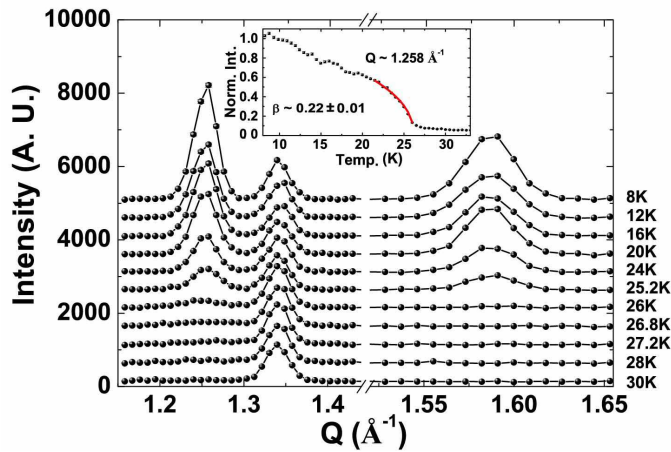}\\
\caption{(Color online) The neutron diffraction data of BaMn$_{2}$Si$_{2}$O$_{7}$ between 1.15 \AA$^{-1}$ and 1.65 \AA$^{-1}$ upon warming. The inset presents the integrated intensity of the magnetic peaks at Q $\sim$ 1.258 \AA$^{-1}$. The results of model (red) fits are the lines overlaying the data points in the inset and the uncertainty bars are derived from counting statistics. The magnetic Bragg peaks of (-2 0 2) and (0 0 2) are found at 1.258 \AA$^{-1}$, and magnetic peaks of (-2 -2 2), (-2 2 2), (0 -2 2), and (0 2 2) are found at 1.586 \AA$^{-1}$.}
  \label{Neutron_HB1A_2}
\end{figure}

The temperature dependence of the phases in the range of 1.15$\le Q \le$1.65 \AA$^{-1}$ were obtained by HB1A, Fig.~\ref{Neutron_HB1A_2}. The magnetic Bragg peaks at 1.258 and 1.586 \AA$^{-1}$ develop at low temperature. The order parameter data at 1.258 \AA$^{-1}$, the inset of Fig.~\ref{Neutron_HB1A_2}, clearly presents that a long-range magnetic ordering transition exists at $T\rm_{N}$=26 K, which is in good agreement with the data of the magnetic susceptibility and specific heat.

Figure~\ref{structure} illustrates the magnetic structure of the Mn$^{2+}$ spins (black arrows). The spin arrangement is antiferromagnetic along the chain ($c$ axis) and ferromagnetic along the $a$ and $b$ directions. Although Mn ions occupy three different positions in this system, there was no evidence suggesting that they have different moment. In addition, the spin components along the $a$-axis were found to be very small from the magnetic structural analysis. Therefore, the spins were aligned along the \emph{b} axis with the ordered moment of 3.9$\mu_B$ for all three Mn$^{2+}$ ions at 4 K.

\subsubsection{Magnetic interactions}
\begin{figure}
 \centering
 \includegraphics[width=0.43\textwidth]{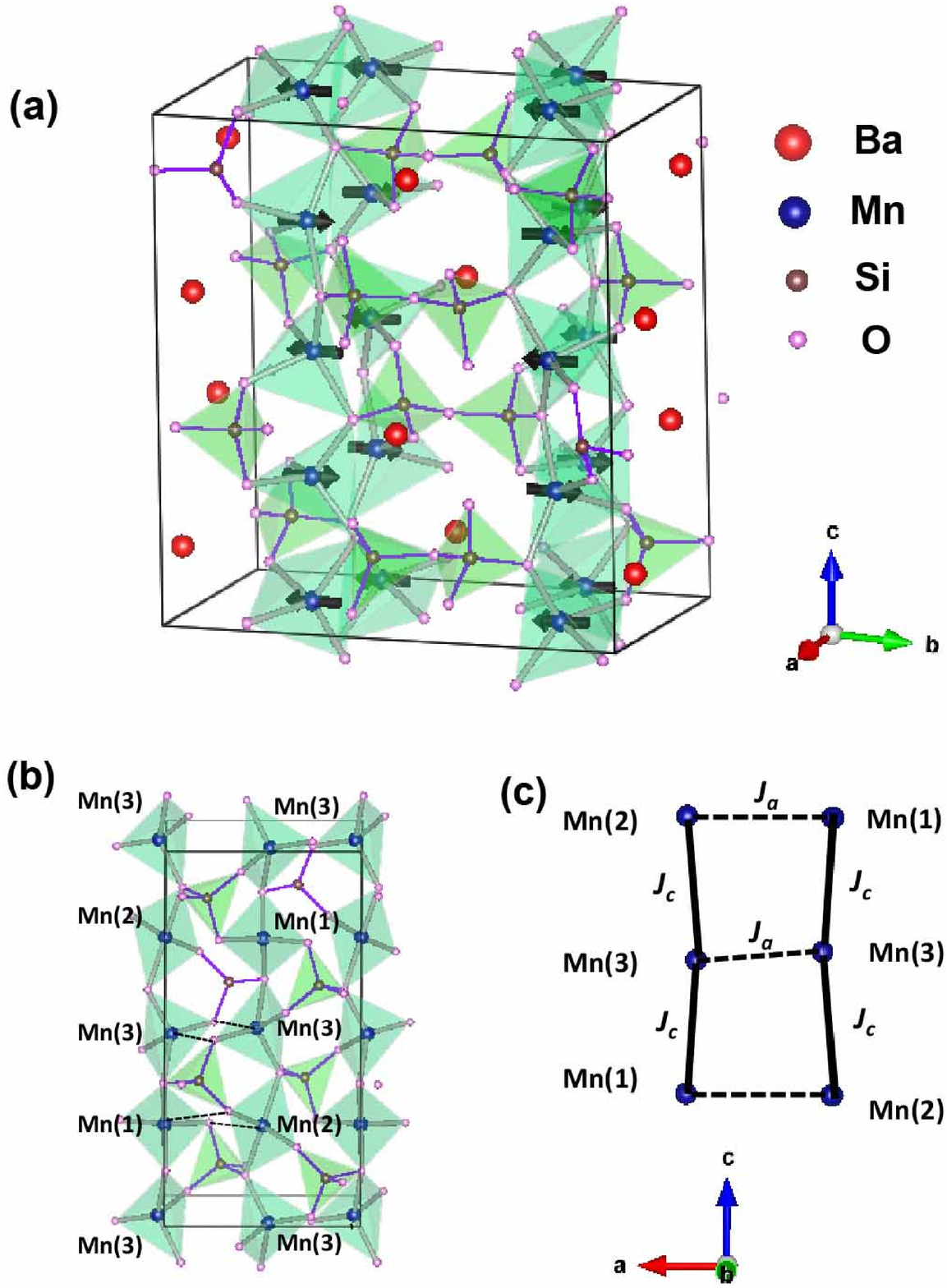}\\
\caption{(Color online) (a) The structure of BaMn$_{2}$Si$_{2}$O$_{7}$: Ba$^{2+}$ (red balls) set along the \emph{c}-axis, Mn$^{2+}$ (blue balls) and Si$^{4+}$ cations (pink balls) are located at the center of O$^{2-}$ anion (cyan balls) tetrahedra, respectively. The Mn-O band is gray, and the Si-O bond is blue. The one-dimensional chain arrangement of MnO$_4$ groups is along the \emph{c} axis. The black arrows present the spins. (b) and (c) show the Mn chain interactions in the \emph{ac} plane. Black dashed lines are the quasi-bonds between Mn$^{2+}$ and O$^{2-}$ anions .}
  \label{structure}
\end{figure}

The Ba atoms occupy the channels formed by these infinite chains and are aligned parallel to the \emph{c} axis. MnO$_4$ tetrahedra are connected with each other via corner-sharing, forming one-dimensional infinite chains along the \emph{c} axis. The Mn-Mn chain is highlighted by connected bonds in gray in Fig.~\ref{structure}, and the distorted zigzag Mn chains are sequenced by -Mn(1)-O(4)-Mn(3)-O(1)-Mn(2)-O(1)-Mn(3)-O(4)- with five distinct Mn-O bonds and two different Mn-O-Mn angles, as shown in Table~\ref{length_angle}.

In BaCu$_{2}$Si$_{2}$O$_{7}$, these corner-shared Cu chains are bridged only by -Si-O- and -O-Si-O-Si-O- bonds from the Si$_2$O$_7$ groups along the \emph{a} and \emph{b} directions, respectively. If we can describe this interchain interaction by the distortion of -SiO$_4$-, it becomes stronger with the substitution of Cu for Mn. (i) Along the \emph{b} axis, the bridge of Cu-chains is -O(1)-Si(1)-O(5)-Si(2)-O(4)-. Although its bond length and angle change significantly with the replacement of Cu for Mn, there are 7 bonds to transfer the interaction between \emph{M} chains, which make the interaction value between the Mn chains along this direction weak; (ii) Along the \emph{a} axis, the -Si-O- bridges connect the \emph{M} chains by 3 bonds, so that the interchain interaction along the $a$ axis is stronger than along \emph{b} axis, although it is still weak. However, the \emph{M}O$_4$ tetrahedron distortion increases the hybridization between $M$ and oxygens, which forms ferromagnetic interaction between \emph{M} moments from different chains and changes the bond lengths and angles between \emph{M} and O ions in different chains as shown with the black dashed lines in Fig.~\ref{structure}(b). This is our proposed mechanism for building up the \emph{transferring bridges} in Ba\emph{M}$_{2}$Si$_{2}$O$_{7}$, which shows how the interaction between the \emph{M} chains is modified. The bond lengths and angles in Ba\emph{M}$_{2}$Si$_{2}$O$_{7}$ are summarized in Table~\ref{length_angle}.

\begin{table*} [tph]
\caption{Selected interatomic distances(\AA) and interatomic bond angles($^\circ$) of Ba\emph{M}$_2$Si$_2$O$_7$ (\emph{M}: Cu, Co, and Mn) at 300 K.}
\begin{tabular}{lccccc}
\hline
\multicolumn{6}{l}{$\, \, \, \, \, \, \, \, \, \, \, \, \, \, \, \, \, \, \, \, \, \, \, \, \, \, \, \, \, \, $BaCu$_2$Si$_2$O$_7^\dag$ $\, \, \, \, \, \, \, \, \, \, \, \, \, \, \, \, \, \, \, \, \, \, \, \, \, \, \, \, \, \, \, \, \, \, \, \, $BaCo$_2$Si$_2$O$_7^\ddag$ $\, \, \, \, \, \, \, \, \, \, \, \, \, \, \, \, \, \, \, \, \, \, \, \, \, \, \, \, \, \, \, \, \, \, \, \, \, \, \, \, \, \, \, $ BaMn$_2$Si$_2$O$_7$} \\
\hline
\multicolumn{6}{l}{along \emph{c}-axis (along \emph{M}-chain)} \\
\multicolumn{6}{l}{$\, \, $ \emph{bond length}} \\
$\,  \,  \,  \,  \,  \,  \,  \,  \,  \,  \,  \,  \,  \,  \,  \,  \,  \,  $Cu-O(4)  & 1.994 & $\,  \, $ Co(1)-O(4) & 1.979 & $\,  \,  $ Mn(1)-O(4)  & 1.930\\
$\,  \,  \,  \,  \,  \,  \,  \,  \,  \,  \,  \,  \,  \,  \,  \,  \,  \,  $Cu-O(4$^\prime$) & 1.995 & $\,  \,  $ Co(2)-O(1) & 2.033 & $\,  \,  $ Mn(2)-O(1)  & 2.125\\
&  & $\,  \,  $ Co(3)-O(1) & 1.971 & $\,  \,  $ Mn(3)-O(1)  & 2.091\\
&  & $\,  \,  $ Co(3)-O(4) & 2.045 & $\,  \,  $ Mn(3)-O(4)  & 2.054\\
\multicolumn{6}{l}{$\, \, \, $ \emph{bond angle}} \\
$\,  \,  \,  \,  \,  \,  \,  \,  $ $\angle$Cu-O(4)-Cu  & 121.04 & $\,  \, $ $\angle$Co(3)-O(4)-Co(1) & 110.47  & $\,  $ $\angle$Mn(3)-O(4)-Mn(1) & 108.67 \\
 & & $\,  \, $ $\angle$Co(3)-O(1)-Co(2) & 127.50 & $\,  $ $\angle$Mn(3)-O(1)-Mn(2)   & 127.27 \\

\multicolumn{6}{l}{along \emph{b}-axis (-Si(1)-O(5)-Si(2)-)} \\
\multicolumn{6}{l}{$\, \, $ \emph{bond length}} \\
$\,  \,  \,  \,  \,  \,  \,  \,  \,  \,  \,  \,  \,  \,  \,  \,  \,  \,  $Si-O(1)  & 1.612 & $\,  \,  $ Si(1)-O(5) & 1.673  & $\,  \,  $ Si(1)-O(5)  & 1.685 \\
&  & $\,  \,  $ Si(2)-O(5) & 1.651  &  $\,  \,  $ Si(2)-O(5)   & 1.610 \\
\multicolumn{6}{l}{$\, \, \, $ \emph{bond angle}} \\
$\,  \,  \,  \,  \,  \,  \,  \,  $ $\angle$Si-O(1)-Si  & 143.72 & $\,  \, $ $\angle$Si(1)-O(5)-Si(2) & 124.47  & $\,  $ $\angle$Si(1)-O(5)-Si(2) & 123.61\\
\multicolumn{6}{l}{along \emph{a}-axis (between \emph{M}-chains)} \\
\multicolumn{6}{l}{$\, \, $ \emph{bond length}} \\
$\,  \,  \,  \,  \,  \,  \,  \,  \,  \,  \,  \,  \,  \,  \,  \,  \,  \,  $Cu-O(2)  & 3.319  & $\,  \,  $ Co$^\prime$(1)-O(2)  & $\, $ 3.173 & $\,  \,  $ Mn$^\prime$(1)-O(2)  & 3.072 \\
$\,  \,  \,  \,  \,  \,  \,  \,  \,  \,  \,  \,  \,  \,  \,  \,  \,  \,  $Cu$^\prime$-O(2)  & 1.930  & $\,  \,  $ Co(2)-O(2)  & $\, $ 2.018 & $\,  \,  $ Mn(2)-O(2)  & 2.231 \\
$\,  \,  \,  \,  \,  \,  \,  \,  \,  \,  \,  \,  \,  \,  \,  \,  \,  \,  $Cu-O(3)  & 1.937  & $\,  \,  $ Co$^\prime$(1)-O(6)  & $\, $ 1.981 & $\,  \,  $ Mn$^\prime$(1)-O(6)  & 1.995 \\
$\,  \,  \,  \,  \,  \,  \,  \,  \,  \,  \,  \,  \,  \,  \,  \,  \,  \,  $Cu$^\prime$-O(3)  & 2.879  & $\,  \,  $ Co(2)-O(6)  & $\, $ 2.633 & $\,  \,  $ Mn(2)-O(6)  & 2.505 \\
&& $\,  \,  $ Co(3)-O(3)  & $\, $ 2.004 & $\,  \,  $ Mn(3)-O(3)  & 2.213 \\
&& $\,  \,  $ Co$^\prime$(3)-O(3)  & $\, $ 2.345 & $\,  \,  $ Mn$^\prime$(3)-O(3)  & 2.288 \\
\multicolumn{6}{l}{$\, \, \, $ \emph{bond angle}} \\
$\,  \,  \,  \,  \,  \,  \,  \,  $ $\angle$Cu-O(2)-Cu$^\prime$ & 78.16 & $\, \, $ $\angle$Co(2)-O(2)-Co$^\prime$(1) & 86.16 & $\,  $ $\angle$Mn(2)-O(2)-Mn$^\prime$(1) & 85.54 \\
$\,  \,  \,  \,  \,  \,  \,  \,  $ $\angle$Cu-O(3)-Cu$^\prime$ & 90.38 & $\, \, $ $\angle$Co(2)-O(6)-Co$^\prime$(1) & 103.48 & $\,  $ $\angle$Mn(2)-O(6)-Mn$^\prime$(1) & 108.02 \\
&& $\, \, $ $\angle$Co(3)-O(3)-Co$^\prime$(3) & 94.39 & $\,  $ $\angle$Mn(3)-O(3)-Mn$^\prime$(3) & 92.68 \\

\multicolumn{6}{l}{distorted SiO$_4$ tetrahedra} \\
\multicolumn{6}{l}{$\, \, $ \emph{bond length}} \\
$\,  \,  \,  \,  \,  \,  \,  \,  \,  \,  \,  \,  \,  \,  \,  \,  \,  \,  $Si-O(2)  & 1.543 & $\,  \, $ Si(1)-O(1) & 1.622 & $\,  \,  $ Si(1)-O(1)   & 1.624 \\
$\,  \,  \,  \,  \,  \,  \,  \,  \,  \,  \,  \,  \,  \,  \,  \,  \,  \,  $Si-O(3)  & 1.562 & $\,  \, $ Si(1)-O(6) & 1.628 & $\,  \,  $ Si(1)-O(6)   & 1.665 \\
$\,  \,  \,  \,  \,  \,  \,  \,  \,  \,  \,  \,  \,  \,  \,  \,  \,  \,  $Si-O(4)  & 1.562 & $\,  \, $ Si(1)-O(7) & 1.614 & $\,  \,  $ Si(1)-O(7)   & 1.621 \\
&  & $\,  \, $ Si(2)-O(2) & 1.605 & $\,  \,  $ Si(2)-O(2)   & 1.549 \\
&  & $\,  \, $ Si(2)-O(3) & 1.617 & $\,  \,  $ Si(2)-O(3)   & 1.599 \\
&  & $\,  \, $ Si(2)-O(4) & 1.638 & $\,  \,  $ Si(2)-O(4)   & 1.733 \\
\multicolumn{6}{l}{$\, \, \, $ \emph{bond angle}} \\
$\,  \,  \,  \,  \,  \,  \,  \,  $ $\angle$O(1)-Si-O(2)  & 103.68 & $\,  \, $ $\angle$O(7)-Si(1)-O(1) & 114.77 & $\,  $ $\angle$O(7)-Si(1)-O(1) &  115.15 \\
$\,  \,  \,  \,  \,  \,  \,  \,  $ $\angle$O(1)-Si-O(3)  & 105.47 & $\,  \, $ $\angle$O(7)-Si(1)-O(6) & 114.27 & $\,  $ $\angle$O(7)-Si(1)-O(6) &  124.70 \\
$\,  \,  \,  \,  \,  \,  \,  \,  $ $\angle$O(1)-Si-O(4)  & 107.80 & $\,  \, $ $\angle$O(7)-Si(1)-O(5) & 109.96 & $\,  $ $\angle$O(7)-Si(1)-O(5) &  104.25  \\
$\,  \,  \,  \,  \,  \,  \,  \,  $ $\angle$O(2)-Si-O(3)  & 112.54 & $\,  \, $ $\angle$O(1)-Si(1)-O(6) & 104.95 & $\,  $ $\angle$O(1)-Si(1)-O(6) &  99.70  \\
$\,  \,  \,  \,  \,  \,  \,  \,  $ $\angle$O(2)-Si-O(4)  & 113.04 & $\,  \, $ $\angle$O(1)-Si(1)-O(5) & 108.39 & $\,  $ $\angle$O(1)-Si(1)-O(5) &  104.36 \\
$\,  \,  \,  \,  \,  \,  \,  \,  $ $\angle$O(3)-Si-O(4)  & 113.41 & $\,  \, $ $\angle$O(6)-Si(1)-O(5) & 109.96 & $\,  $ $\angle$O(6)-Si(1)-O(5) &  106.97 \\
\hline
\multicolumn{6}{l}{\dag$\, $ are calculated from reference [], \ddag $\, $ are calculated from reference [17]. \emph{M} and \emph{M}$^\prime$ are}\\
\multicolumn{6}{l}{ in different chains.}\\
\end{tabular}
\label{length_angle}
\end{table*}

\begin{table} [tph]
\caption{Selected interatomic distances(\AA) and bond angle($^\circ$) of BaMn$_2$Si$_2$O$_7$ at 4 K. }
\begin{tabular}{cclc}
\hline
\multicolumn{4}{l}{along \emph{c}-axis (along \emph{M}-chain)} \\
Mn(1)-O(4)  & $\, $ 2.091 & $\, \, $ $\angle$Mn(1)-O(4)-Mn(3) & $\, $ 103.37\\
Mn(2)-O(1)  & $\, $ 2.216 & $\, \, $ $\angle$O(4)-Mn(3)-O(1) & $\, $ 110.68\\
Mn(3)-O(1)  & $\, $ 2.083 & $\, \, $ $\angle$Mn(3)-O(1)-Mn(2) & $\, $ 127.37\\
Mn(3)-O(4)  & $\, $ 1.951 & $\, \, $ $\angle$O(1)-Mn(2)-O(1) & $\, $ 160.89\\
& & $\, \, $ $\angle$O(4)-Mn(1)-O(4) & $\, $ 111.92\\
\multicolumn{4}{l}{along \emph{b}-axis (-Si(1)-O(5)-Si(2)-)}\\
Si(1)-O(5)  & $\, $ 1.635 & $\, \, $ $\angle$Si(1)-O(5)-Si(2) & $\, $ 116.18\\
Si(2)-O(5)  & $\, $ 1.571 &&\\
\multicolumn{4}{l}{along \emph{a}-axis (between \emph{M}-chains)}\\
Mn$^\prime$(1)-O(2)  & $\, $ 3.097 & $\, \, $ $\angle$Mn(2)-O(2)-Mn$^\prime$(1) & $\, $ 84.11\\
Mn(2)-O(2)  & $\, $ 2.260 &&\\
Mn$^\prime$(1)-O(6)  & $\, $ 2.082 & $\, \, $ $\angle$Mn(2)-O(6)-Mn$^\prime$(1) & $\, $ 108.49\\
Mn(2)-O(6)  & $\, $ 2.400 &&\\
Mn(3)-O(3)  & $\, $ 2.061 & $\, \, $ $\angle$Mn(3)-O(3)-Mn$^\prime$(3) & $\, $ 90.99\\
Mn$^\prime$(3)-O(3)  & $\, $ 2.276 &&\\
\multicolumn{4}{l}{distorted SiO$_4$ tetrahedra} \\
Si(1)-O(1)   & $\, $ 1.747 & $\, \, $ $\angle$O(1)-Si(1)-O(5)& $\, $ 104.66\\
Si(1)-O(6)   & $\, $ 1.563 & $\, \, $ $\angle$O(1)-Si(1)-O(7)& $\, $ 107.47\\
Si(1)-O(7)   & $\, $ 1.588 & $\, \, $ $\angle$O(1)-Si(1)-O(6)& $\, $ 100.69\\
& & $\, \, $ $\angle$O(5)-Si(1)-O(6)& $\, $ 108.53\\
& & $\, \, $ $\angle$O(5)-Si(1)-O(7)& $\, $ 108.50\\
& & $\, \, $ $\angle$O(6)-Si(1)-O(7)& $\, $ 125.04\\
Si(2)-O(2)   & $\, $ 1.523 & $\, \, $ $\angle$O(4)-Si(2)-O(5)& $\, $ 104.72\\
Si(2)-O(3)   & $\, $ 1.714 & $\, \, $ $\angle$O(2)-Si(2)-O(4)& $\, $ 116.54\\
Si(2)-O(4)   & $\, $ 1.719 & $\, \, $ $\angle$O(3)-Si(2)-O(4)& $\, $ 90.04\\
& & $\, \, $ $\angle$O(2)-Si(2)-O(3)& $\, $ 115.37\\
& & $\, \, $ $\angle$O(2)-Si(2)-O(5)& $\, $ 116.54\\
& & $\, \, $ $\angle$O(3)-Si(2)-O(5)& $\, $ 118.60\\
\hline
\multicolumn{4}{l}{\emph{M} and \emph{M}$^\prime$ are belong to different chains in ${ac}$ plan.}\\
\end{tabular}
\label{length4K}
\end{table}

\section{Discussion}
In order to understand the critical behavior of the magnetic sublattice for BaMn$_{2}$Si$_{2}$O$_{7}$, the temperature dependence of the magnetic peak was analyzed, Fig.~\ref{Neutron_HB1A_2}(inset). A simple power law was applied to fit the integrated intensities, I, of the magnetic diffraction peaks near the transition temperature,
\begin{equation}
I  = I_0 ( 1 - \frac{T}{T_{N}} )^{2\beta} \, ,
\label{fitI}
\end{equation}
where \emph{T}$_N$ is the Ne\'{e}l temperature, \emph{I}$_0$ is the integrated intensity at base temperature, and $\beta$ is the order parameter critical exponent.

The obtained critical exponent is 0.22 $\pm$ 0.1 for the magnetic peak, shown in Fig.~\ref{Neutron_HB1A_2}, which is incompatible with the three-dimensional Heisenberg model ($\beta$ $\sim$ 0.345), and a typical value for a low dimensional material.

In the Ba\emph{M}$_2$Si$_2$O$_7$ system, the $M$ chain is formed by superexchange via the oxygen bridges, which makes the interactions of the -Mn-O-Mn- chain in BaMn$_2$Si$_2$O$_7$ more complicated than the -Cu-O-Cu- chain in BaCu$_2$Si$_2$O$_7$. There are three different positions for the Mn$^{2+}$ ions in BaMn$_2$Si$_2$O$_7$, so it is hard to compare the superexchange interactions directly by single bond angles and lengths. The superexchange between 3$d$-2$p$-3$d$ electrons can be described by the overlap integral \emph{b}$^2$, and \emph{T}$\rm_N$ $\sim$ \emph{b}$^2$/\emph{U}, where \emph{U} is the interatomic Coulomb energy. Both bond-length modulation and angles of the \emph{M}-O-\emph{M}$^\prime$ bond with an angle of 180$^\circ$-$\omega$ are incorporated in the expression for the overlap integral given by, \cite{JSZhou}
\begin{equation}
  b^2 \sim {\rm cos}^4 (\omega/2)/d^7 \, ,
\label{fiteq}
\end{equation}
where \emph{d} is the bond length.

Table~\ref{b2} shows the overlap integrals along different directions of Ba\emph{M}$_{2}$Si$_{2}$O$_{7}$ (\emph{M}: Cu, Co, and Mn) at 300 K and BaMn$_{2}$Si$_{2}$O$_{7}$ at 4 K. Along the \emph{c} axis, where the \emph{M}-O-\emph{M} chain is formed, the interaction between the \emph{M} ions becomes weaker as the $M$ site is changed from Cu to Co and from Co to Mn due to the different environment surrounding the \emph{M}$^{2+}$ ion. In BaCu$_{2}$Si$_{2}$O$_{7}$, Cu$^{2+}$ is located at the center of a square planar with a small distortion. The only magnetic orbital of the Cu$^{2+}$ is 3\emph{d}$_{x^2-y^2}$, which points along the $c$ axis, therefore, the superexchange interaction via Cu-O(4)-Cu bond is strong \cite{Hayn}. In BaCo$_{2}$Si$_{2}$O$_{7}$, the square planar of CoO$_4$ is distorted to form a tetrahedron, which leads the Co$^{2+}$ orbitals to split into \emph{e}$_g$ and \emph{t}$_{2g}$ with \emph{e}$_{g}$ orbitals fully occupied and \emph{t}$_{2g}$ orbitals half occupied. Since the three \emph{t}$_{2g}$ orbitals do not point along the neighboring oxygen ions, the hybridization between the Co$^{2+}$ and oxygen orbitals is not large. The similar crystal field effect due to the tetrahedral distortion happens in BaMn$_{2}$Si$_{2}$O$_{7}$. Although both the \emph{e}$_{g}$ and \emph{t}$_{2g}$ orbitals are half occupied in Mn$^{2+}$, the cation size-effect is more significant so that the superexchange interaction is considered to be even weaker. In addition, the MnO$_4$ tetrahedron is more distorted at lower temperatures, therefore the intrachain interaction becomes weaker as the temperature decreases.

Along the \emph{b} axis, the the overlap of -Si-O-Si- at 300 K is strong in the all three samples due to the 2$p$-2$p$ electrons interaction between Si$^{4+}$ and O$^{2-}$ ions, but the correlation between the chains are transferred by 5 atoms and 7 bonds and the interactions are not expected to be strong.

Along the \emph{a} axis, there are two mechanisms to describe the interaction between the \emph{M} chains: 1) the \emph{M} chain is connected by the SiO$_4$ tetrahedra. Although the distance between \emph{M} chains along the $a$ axis is not large, the interactions are transferred by 3 atoms and 5 bonds, which makes the interaction weak. 2) The hybridization effect between \emph{M} and O ions is noticeable enough to increase the linking between the chains, as shown in Table~\ref{b2}. For example, if the bonds of Mn(2)-O(1) and Mn(3)-O(6) in BaMn$_{2}$Si$_{2}$O$_{7}$ could be formed, as described above, which occupies the cross-corner oxygens between MnO$_{4}$ and SiO$_{4}$, the average overlap integral is calculated to be around 0.0014 and close to the values along the chain. Therefore, the interaction should not be ignored. However, in BaCu$_{2}$Si$_{2}$O$_{7}$ both \emph{b}$^2$'s of Cu-O(2)-Cu$^\prime$ and Cu-O(3)-Cu$^\prime$ are $\sim$7\% of the value of Cu-O(4)-Cu (along the chain), which makes BaCu$_{2}$Si$_{2}$O$_{7}$ more one-dimensional.

\begin{table} [tph]
\caption{The overlap integral(\emph{b}$^2$) of Ba\emph{M}$_2$Si$_2$O$_7$ (\emph{M}: Cu, and Co) ($^\circ$) at 300\emph{K} and BaMn$_2$Si$_2$O$_7$ at 4 and 300\emph{K}. }
\begin{tabular}{cccc}
\hline
$\, \, \, \, $BaCu$_2$Si$_2$O$_7$ & $\, \, \, \, $BaCo$_2$Si$_2$O$_7$ & \multicolumn{2}{c}{$\, \, \, \, $ BaMn$_2$Si$_2$O$_7$ $\, \, \, \, $} \\
\cline{3-4}
\multicolumn{4}{c}{$\, \, \, \, \, \, \, \, \, \, \, \, \, \, \, \, \, \, \, \, \, \, \, \, \, \, \, \, \, \, \, \, \, \, \, \, \, \, \, \, \, \, \, \, \, \, \, \, \, \, \, \, \, \, \, \, \, \, \, \, \, \, \, \, \, \, \, \, $ 300 K $\, \, \, \, \, $ 4 K} \\
\hline
\multicolumn{4}{l}{along \emph{c}-axis (along \emph{M}-O-\emph{M} chain)} \\
0.0046 & 0.0034 & 0.0035 & 0.0028 \\
  --   & 0.0051 & 0.0034 & 0.0030 \\
\multicolumn{4}{l}{along \emph{a}-axis (between \emph{M}-O-\emph{M} chain)} \\
0.0002 & 0.0003 & 0.0002 & 0.0002 \\
0.0005 & 0.0011 & 0.0015 & 0.0015 \\
-- & 0.0013 & 0.0009 & 0.0011 \\
\hline
\end{tabular}
\label{b2}
\end{table}

Therefore, in BaCu$_{2}$Si$_{2}$O$_{7}$, the interchain coupling along the \emph{a} axis (\emph{J}$_a$) should not be ignored compared to that along the \emph{c} axis (\emph{J}$_c$) as Eq.~\ref{Heisen}. The more appropriate Hamiltonian would be expressed as,
\begin{equation}
H = - 2J_a\sum_{i, j \parallel a} {\bf S_i} \cdot {\bf S_j} - 2J_c\sum_{i, j \parallel c} {\bf S_i} \cdot {\bf S_j} \, ,
\label{Heisen_ac}
\end{equation}
where ${\bf \emph{S}}_i$ and ${\bf \emph{S}}_j$ are the spin vectors on the \emph{i}th and \emph{j}th sites. The subscripts $i$, $j$ $\parallel$ \emph{a} and \emph{c} indicate that sums are restricted to nearest-neighbor spins along the \emph{a} and \emph{c} axes. Exchange interactions are restricted to nearest neighbor spins along the \emph{a} and \emph{c} axes.  Positive and negative \emph{J} represent ferromagnetic and antiferromagnetic exchange interactions, respectively. The spin structure shown above suggests that \emph{J}$_c$ is antiferromagnetic and \emph{J}$_a$ is ferromagnetic.

From mean-field theory, the exchange energy could be expressed as,
\begin{equation}
J_a + \mid J_c \mid = \frac{3 k_B T_N}{2S(S+1)} \, ,
\label{exchange}
\end{equation}
which agrees with the report of Schulz \emph{et al}. \cite{Schulz}

Since $\mid$\emph{J}$\mid$ $\propto$ \emph{b}$^2$, $\mid$\emph{J}$_c$$\mid$ $\approx$ 2\emph{J}$_a$ from the values of overlap integral along different directions, as shown in Table~\ref{b2}. As \emph{S}=5/2 and \emph{T}$\rm_N$=26 K, \emph{J}$_c$ is estimated to be - 3 K from Eq.~\ref{exchange}, which is comparable to the estimated value from the magnetic susceptibility data, -6 K. An inelastic neutron scattering experiment using a single crystal is desired to determine the magnetic interactions unambiguously.

\section{Conclusion}

Neutron powder diffraction, magnetic susceptibility, and heat capacity techniques are applied to study the structural and magnetic properties of BaMn$_{2}$Si$_{2}$O$_{7}$, which is considered to be a quasi-one-dimensional antiferromagnet.
The magnetic ground state is determined to be a three dimensional antiferromagnetic ordered phase. $T\rm_N$ (26 K) is rather high, compared to the magnetic coupling along the chain (-6 K) and the large spin value (S = 5/2). We found from the structural analysis that the high transition temperature is ascribed to the nonnegligible interchain coupling along the $a$ axis.

\begin{center}
$\textbf{ACKNOWLEDGEMENT}$
\end{center}

The research at Oak Ridge National Laboratory was sponsored by the Scientific User Facilities Division (JM, MM, CDDC, TH, WT, AAA, SXC) and Materials Sciences and Engineering Division (JQY), Office of Basic Energy Sciences, US Department of Energy.

\bibliographystyle{apsrev}

\end{document}